\documentclass{webofc}
\usepackage[varg]{txfonts}   
\usepackage{listings}
\usepackage{hyperref}
\newcommand{\rrl}{RR~Lyr\ae}

\begin{document}
\title{Observations of Field and Cluster \rrl{} with Spitzer}
\subtitle{Towards High Precision Distances with Population II Stellar Tracers}

\author{
\firstname{Massimo} \lastname{Marengo} \inst{1} \and
\firstname{Jillian} \lastname{Neeley} \inst{1} \and
\firstname{Giuseppe} \lastname{Bono} \inst{2,3} \and
\firstname{Vittorio} \lastname{Braga} \inst{2,3,4} \and
\firstname{Massimo} \lastname{Dall'Ora} \inst{5} \and
\firstname{Marcella} \lastname{Marconi} \inst{5} \and
\firstname{Nicolas} \lastname{Trueba} \inst{1} \and
\firstname{Davide} \lastname{Magurno} \inst{2,3} \and
\firstname{} \lastname{the CRRP Collaboration} \inst{}
}

\institute{
Department of Physics and Astronomy, Iowa State University, Ames, IA 50010, USA 
\and
Department of Physics, Universit\`a di Roma Tor Vergara, via della Ricerca Scientifica 1, I-00133 Roma, Italy           
\and
INAF-Osservatorio Astronomico di Roma, via Frascati 33, I-00040 Monte Porzio Catone, Italy
\and
Departamento de F\'isica, Facultad de Ciencias Exactas, Universidad Andr\'es Bello, Santiago, Chile
\and
INAF-Osservatorio Astronomico di Capodimonte, Salita Moiarello 16, I-80131 Napoli, Italy
}

\abstract{%
  We present our project to calibrate the \rrl{} period-luminosity-metallicity relation using a sample of Galactic  calibrators in the halo and globular clusters.
}
\maketitle

\section{Introduction}\label{sec:intro}

The availability of reliable, high-precision, \rrl{} distances could be a game-changer not just for tracing old stellar populations in the Milky Way and other Local Group galaxies, but also for renewed efforts to establish an independent cosmological distance scale entirely based on Population II indicators \cite{2016ApJ...832..210B}. This will be crucial to resolve the current $\ge 3 \sigma$ tension existing between different measurements of the Hubble constant and other cosmological parameters. Measurements obtained with methods based on Population I stellar indicators, anchored on Cepheids, are at odd with determinations based on the Cosmic Background Radiation and Baryonic Acoustic Oscillations \cite{2016ApJ...826...56R}. If confirmed, this tension could be an indication of unknown physics, missing in the current formulation of $\Lambda$CDM cosmological models.

\begin{figure}
\centering
\sidecaption
\includegraphics[width=7.5cm,clip]{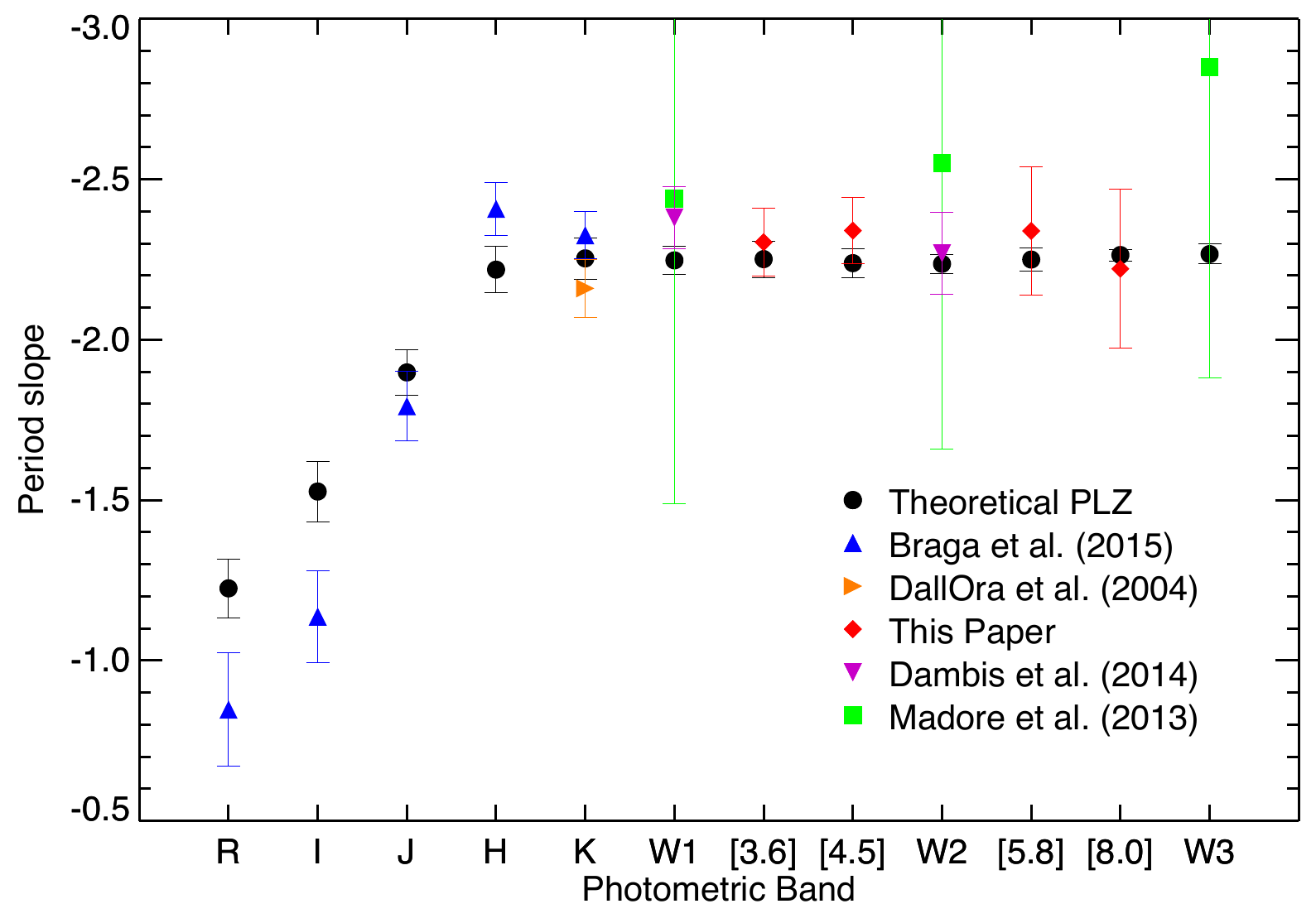}
\caption{Period slope of \rrl{} PLZ relations in optical and infrared bands. Different symbols indicate slope observational values from cluster and field \rrl{}, compared with the theoretical prediction presented in \cite{2015ApJ...808...50M} and \cite{2017ApJ...XXX.....N} (black circles). Note that the slope is very shallow in the visible, and then increases in the NIR up to the $H$ band. For wavelengths longer than $\sim 2$~$\mu$m the period slope becomes constant, converging to the value expected for brightness variations entirely driven by the change in radius.}
\label{fig:PLslope}       
\end{figure}

Historically, the adoption of \rrl{} as precision standard candles has been limited by the lack of a proper period-luminosity (PL) relation. Their visible magnitude only shows a shallow dependence from their pulsation period, and the magnitude vs. metallicity relation commonly used to derive their photometric parallaxes is characterized by a  large intrinsic scatter ($> 5$\%). The situation however changes at near- and mid-infrared wavelengths (NIR and MIR thereafter), for three main reasons: (i) in the NIR the slope in the \rrl{} PL relation steadily increases (see Figure~\ref{fig:PLslope}); (ii) the intrinsic scatter due to evolutionary effects and temperature dependence is greatly reduced in the infrared, because of the narrowing of the instability strip width, and because in the MIR the brightness variations are mainly driven by the radius changes; (iii) in the NIR the effect of reddening is one order of magnitude less than in the visible, with an extra factor of 3 reduction in the MIR \cite{2016MmSAIt.XXX..XXXB}. These predictions were recently confirmed, observationally, both in NIR \cite{2004ApJ...610..269D} and MIR \cite{2011ApJ...738..185K, 2013ApJ...776..135M} bands.

With the upcoming availability of large aperture infrared space telescopes (JWST in the NIR and MIR, and WFIRST that will operate at wavelengths up to 2~$\mu$m), it is time to provide a robust calibration of \rrl{} PL relations, and characterize their dependence from metallicity. This paper presents a summary of our calibration efforts based on the MIR observation of Galactic \rrl{} with the Spitzer Space Telescope InfraRed Array Camera (IRAC) \cite{2004ApJS..154...10F}, part of the Carnegie \rrl{} Program (CRRP), as well as ground-based observations at visible and NIR wavelengths.

\section{Calibrating the Period-Luminosity-Metallicity (PLZ) Relations}\label{sec:PLZ}

The absolute brightness of \rrl{} can be approximated by a linearized expression of their pulsation period $P$ and iron content [Fe/H] (as proxy for their detailed abundances):

\begin{equation}
M_\lambda = a_\lambda + b_\lambda \cdot \log P + c_\lambda \cdot \textrm{[Fe/H]}
\label{eq:PLZ}
\end{equation}

\noindent
where the dependence on metallicity has been factorized in the $c_\lambda$ term. Detailed time-dependent hydrodynamic models (such as \cite{2015ApJ...808...50M}, see also M. Marconi contribution to this volume) show that the metallicity slope $c_\lambda$ is only weakly dependent from wavelength, since the metallicity dependence of these stars' bolometric corrections effectively cancels-out the corresponding changes in effective temperature, resulting in a minimal change in color as [Fe/H] is increased. Furthermore, in the MIR the stellar photosphere is well described by the Rayleigh-Jeans approximation, and lacks significant spectral features. For this reasons the PLZ metallicity slope approaches a constant value $c$. In this formulation, the period slope $b_\lambda$ is assumed independent from metallicity, a result supported by models; in the MIR this term converges to the value expected for brightness variations driven by changes in radius ($b_\lambda \simeq - 2.3$ for fundamental mode pulsators, see Figure~\ref{fig:PLslope}).

To be truly useful, \rrl{} PLZs need however to be calibrated observationally. This can be achieved by observing a number of stars in the field and globular clusters (GCs) with known true distance modulus $\mu_i$, having a broad range of periods $P_i$ and metallicity [Fe/H]$_i$. The average magnitudes $m_{\lambda,i}$ of each of these calibrators need to be measured at multiple wavelengths, in order to characterize their individual extinction $A_{V,i}$ (after assuming an appropriate reddening law $[A_\lambda/A_V]$), so that:

\begin{equation}
m_{\lambda,i} - \mu_i = a_\lambda + b_\lambda \cdot \log P_i + c_\lambda \cdot \textrm{[Fe/H]}_i + A_{V,i} \cdot [A_\lambda/A_V]
\label{eq:PLZ-Av}
\end{equation}

Equation~\ref{eq:PLZ-Av} can then be fit to derive the PLZ coefficients $a_\lambda$, $b_\lambda$ and $c_\lambda$ (and the nuisance parameter $A_{V,i}$ for each star) for each band in which the calibrators have been observed. To minimize the uncertainty in the best fit PLZ coefficients, the sample of \rrl{} calibrators in our program have been selected so that they will have Gaia parallax determined better than $\sim 1$\%. For all stars we are also acquiring a homogeneous set of metallicity (with precision better than $\sim 0.1$~dex), using optical and NIR high resolution spectroscopic observations.

\newpage

\begin{figure*}
\centering
\includegraphics[width=0.86\textwidth,clip]{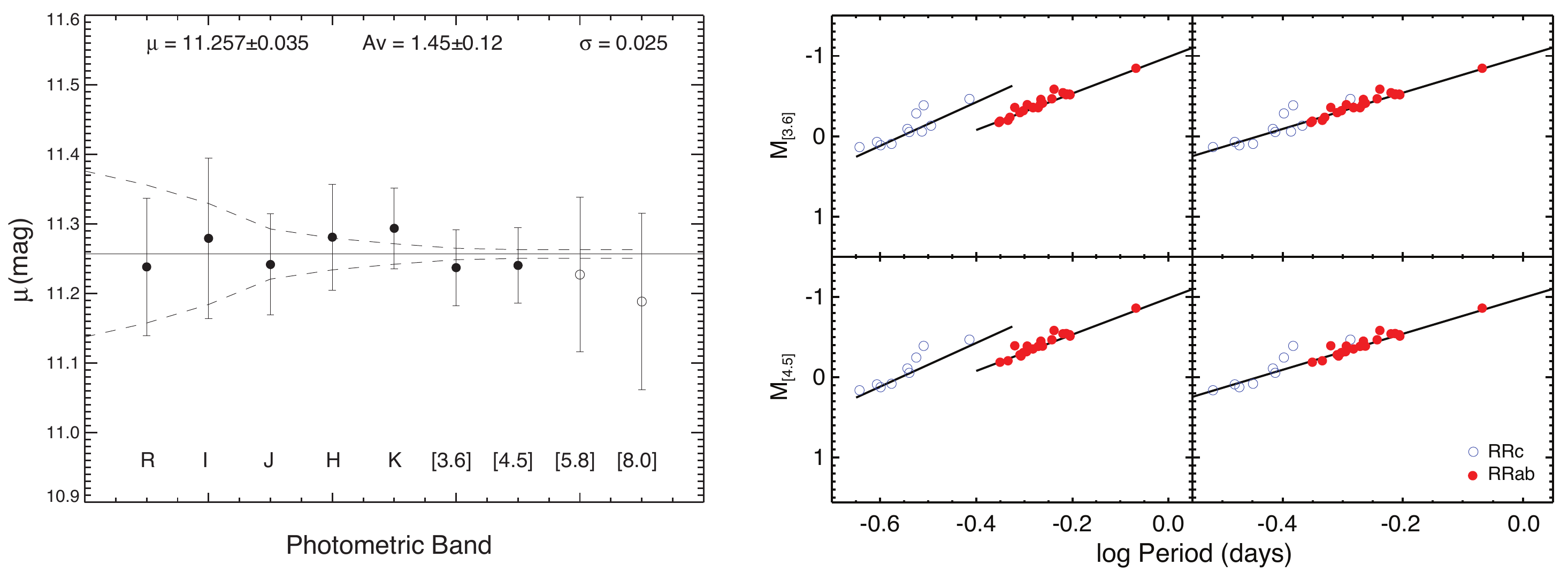}
\caption{\emph{Left} --- Multi-wavelength best fit distance modulus and visual extinction for the Galactic globular cluster M4 (NGC~6121). \rrl{} stars in the cluster have been fitted using the theoretical PLZ relations in \cite{2015ApJ...808...50M}, for the [Fe/H] value of the cluster. The two longest wavelength IRAC bands (single-epoch archival observations) have not been included in the fit. The IRAC 3.6 and 4.5~$\mu$m bands provide the strongest constraints for the distance modulus, while the optical $R$ and $I$ bands are essential to fit $A_V$, based on the adopted extinction law. \emph{Right} --- PL relations for first overtone (RRc, blue empty circles) and fundamental mode (RRab, red filled circles) M4 \rrl{} at 3.6 and 4.5~$\mu$m (fundamentalized relations on the right). Both figures are adapted from \cite{2017ApJ...XXX.....N}.}
\label{fig:M4}       
\end{figure*}

\section{Calibrating the Period Slope with Galactic Globular Clusters}\label{sec:clusters}

From Equation~\ref{eq:PLZ-Av} it is clear that even small uncertainties in the observed quantities (especially [Fe/H], rarely known better than $\sim 0.2$~dex), and variations in the reddening law will introduce a large scatter, and possibly a period-dependent bias, in the observed PLZ. This makes it difficult to simultaneously fit all PLZ coefficients, and the extinction $A_V$, using an heterogeneous sample of calibrator \rrl{} located at different distance, and having different metallicity. The period slope $b_\lambda$, however, can be determined independently from all other parameters by observing a sample of single-metallicity GCs. The advantage of using GCs is of course that the distance of the cluster from Earth is much larger than the relative spatial distribution of the individual stars within the GC, and variable extinction in front of the cluster only contributes to a small unbiased scatter around the overall \rrl{} PL: for \rrl{} in a GC the period slope $b_\lambda$ can be accurately fit as the slope of the PL relation constructed with the reddened apparent magnitudes.

Figure~\ref{fig:M4} (adapted from \cite{2017ApJ...XXX.....N}) presents an example based on our observations of M4 (NGC~6121). The right panel shows the PLZ for first overtone and fundamental mode \rrl{}, derived for IRAC 3.6 and 4.5~$\mu$m bands. The period slope we measured in the NIR \cite{2015ApJ...799..165B} and MIR \cite{2017ApJ...XXX.....N} are shown in Figure~\ref{fig:PLslope}, and are in good agreement with their theoretical values. Repeating this measurement for the other single-metallicity clusters observed as part of the CRRP program will allow us to refine the fitted values of $b_\lambda$, and check that this term is indeed independent from [Fe/H]. Once the $b_\lambda$ are determined, we will use our sample of field calibrator \rrl{} to fit the remaining parameters $a_\lambda$ and $c_\lambda$.

The left panel of Figure~\ref{fig:M4} assesses how well, if the PLZ is fully calibrated, it can be used to accurately derive the distance and average reddening of GCs. For this purpose we have used the photometric data of all \rrl{} in M4, at $RIJHK$ and IRAC MIR bands, fitting at each wavelength the cluster reddened distance modulus by assuming the validity of the theoretical PLZ from \cite{2015ApJ...808...50M}. We have then derived the average distance modulus and extinction, adopting the peculiar reddening law for M4 derived by \cite{2012AJ....144...25H}. The result is consistent with independent measurements of the cluster's distance and average $A_V$, and shows that this method is capable of providing accurate distances for individual Galactic GCs with a total uncertainty of just a few percent.

\newpage

\begin{figure*}
\centering
\includegraphics[width=0.89\textwidth,clip]{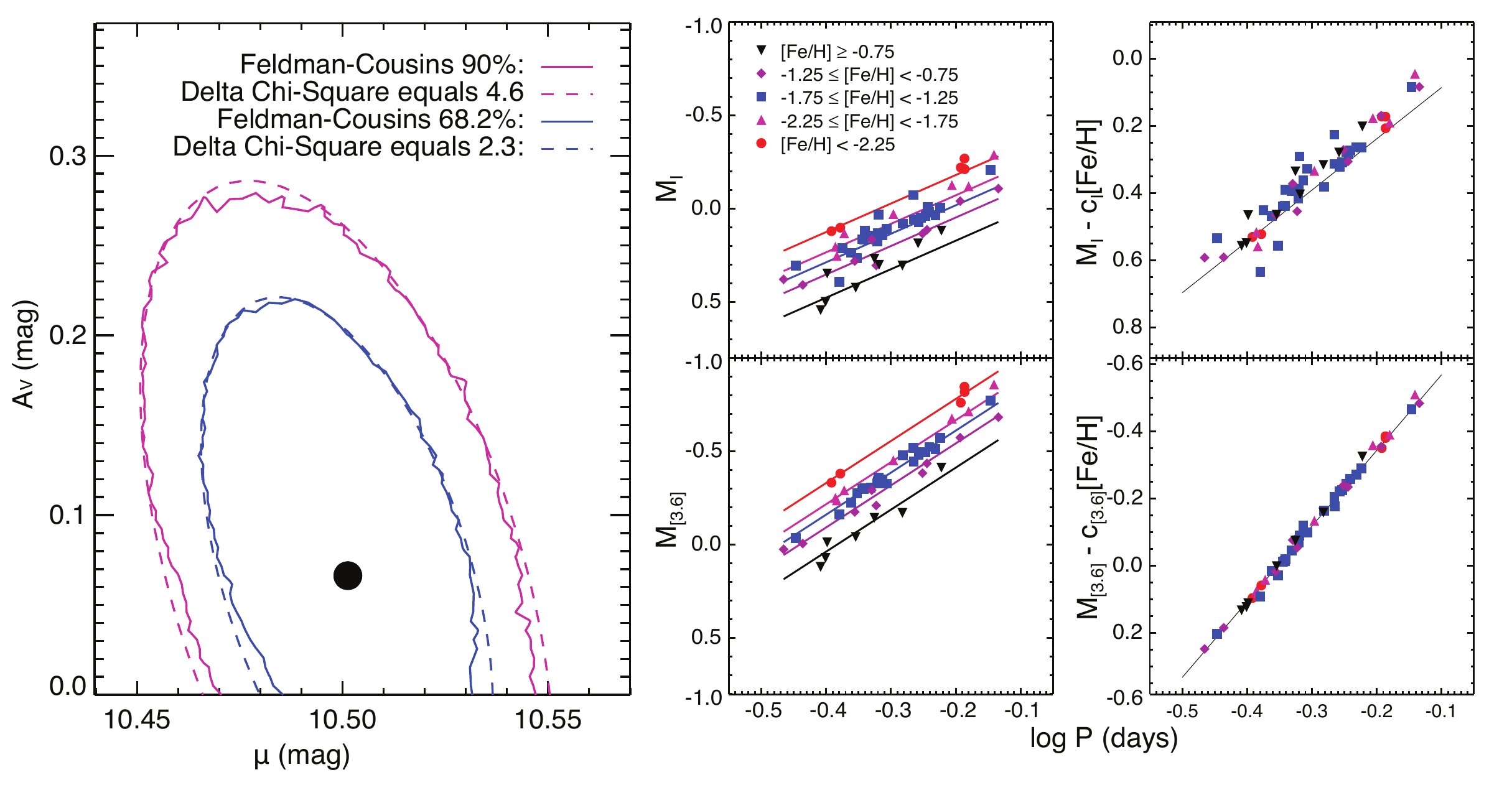}
\caption{\emph{Left} --- Best fit value distance modulus and extinction, 1$\sigma$ and 90\% confidence regions for the field \rrl{} AP~Ser. The fit is based on the theoretical PLZ relations derived in \cite{2015ApJ...808...50M}. Confidence regions calculated with the Feldman and Cousins method \cite{1998PhRvD..57.3873F} are preferred if one of the fitted parameters is bound by physical boundaries (e.g. $A_V \ge 0$). \emph{Center and Right} --- PLZ relations of Galactic field \rrl{} in the $I$ and IRAC 3.6~$\mu$m bands, binned based on their [Fe/H]. The scatter due to the difference in metallicity is $\sim 0.13$ magnitudes over the metallicity range in the Galactic halo. The two panels on the right show how the scatter is much reduced when the metallicity term is subtracted by the stars' absolute magnitude, especially in the MIR band, where it becomes less than 0.02~magnitudes. Adapted from \cite{2017ApJ...XXX.....N}.}
\label{fig:calib}       
\end{figure*}

\section{High Precision Individual \rrl{} Distances}\label{sec:fieldRRL}

The theoretical PLZ relations derived by \cite{2015ApJ...808...50M} can be used to estimate the effect of metallicity on individual \rrl{} distances. This is shown in Figure~\ref{fig:calib}, where the distance modulus and extinction of each calibrator in our sample was fit using the theoretical PLZ relations (one example in the left panel, with the best fit confidence level derived using the Feldman and Cousin method \cite{1998PhRvD..57.3873F}, appropriate when a fitting parameter is bound by a physical boundary, e.g. $A_V \ge 0$). The central panels show the distribution of the calibrators' magnitudes (derived using their best fit distance and extinction) in the $I$ and IRAC 3.6~$\mu$m bands, binned based on their [Fe/H]. In both wavelengths the dispersion is of the order of $\pm 0.13$ magnitudes. Once the metallicity term $c_\lambda \cdot \textrm{[Fe/H]}$ is subtracted from the stars' absolute magnitude, the spread is significantly reduced (right). This is especially true in the MIR bands, where temperature and evolutionary effects are vanishing, leading to drastically smaller intrinsic scatter in the resulting PLZ (once metallicity effects are removed, we found $\sigma \simeq 0.02$~mag in the IRAC bands).

\begin{figure*}
\centering
\includegraphics[width=0.68\textwidth,clip]{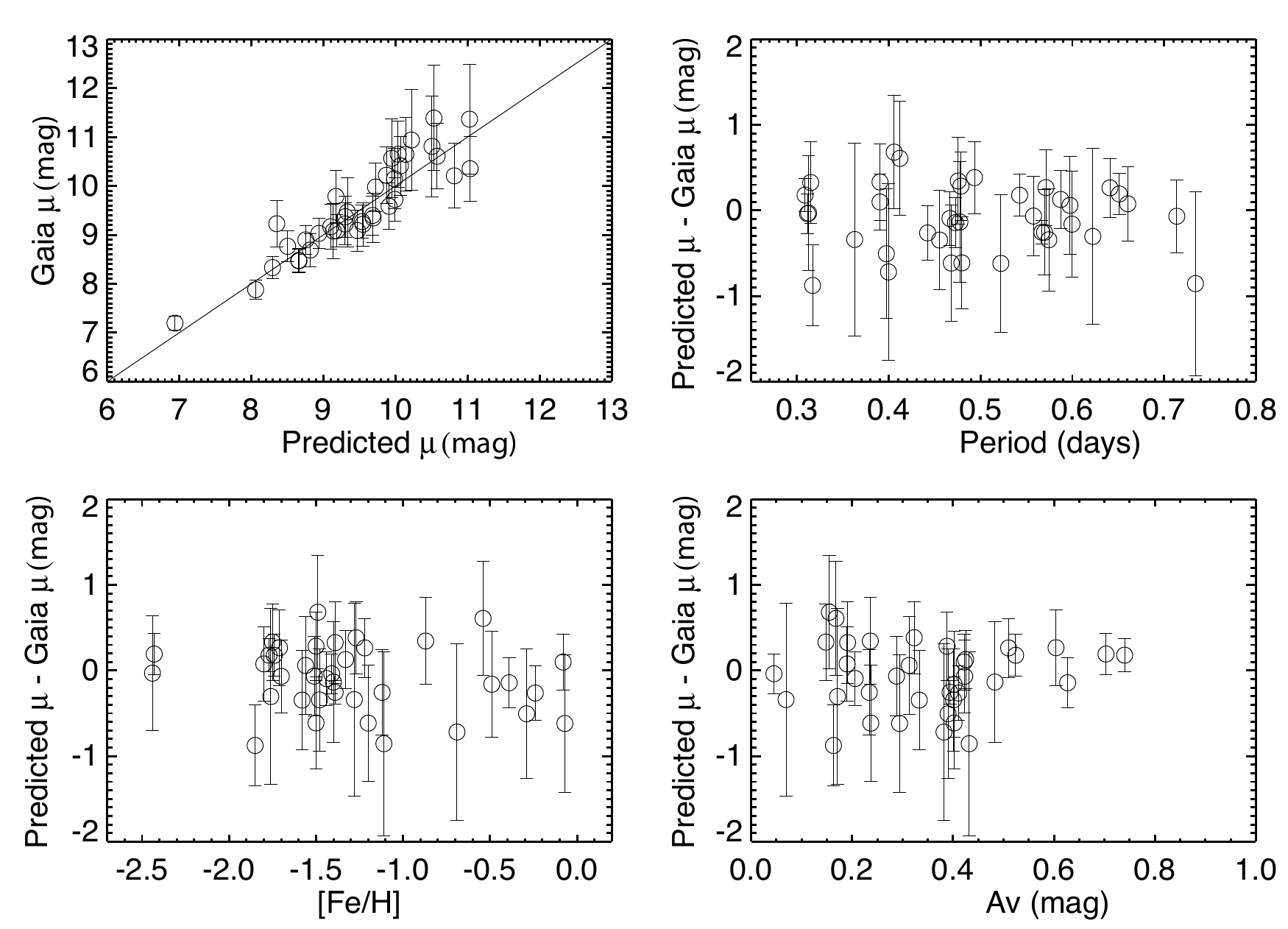}
\caption{Comparison between the best fit distance moduli of Galactic field \rrl{}, obtained using the theoretical PLZ relations described in \cite{2015ApJ...808...50M}, and Gaia DR1 \cite{2016A&A...595A...2G} distances. The two sets of value match better than $1\sigma$, and do not show any bias due to period, metallicity or best fit extinction. Adapted from \cite{2017ApJ...XXX.....N}.}
\label{fig:gaia}       
\end{figure*}

To test the validity of the theoretical PLZ derived by \cite{2015ApJ...808...50M} and \cite{2017ApJ...XXX.....N} we have compared our best fit distance moduli with the Gaia first data release (DR1, \cite{2016A&A...595A...2G}) distances. Figure~\ref{fig:gaia} shows that our results are within $1 \sigma$ of the Gaia measurements, and that the residuals are not biased in either period, metallicity or the value of extinction derived with our best fitting procedure.

A detailed analysis of these results is provided in \cite{2017ApJ...XXX.....N}. Here we want to highlight how precisely calibrating the metallicity term in the \rrl{} PLZ is essential to have reliable distances for individual stars, a necessary condition for using them as accurate tracers of old stellar populations. The typical $\sim 0.2$~dex uncertainty in most spectrophotometric determination of metallicity, just by itself, would result in a distance uncertainty of almost $\sim 4$~\%, stressing the need of accurately measuring the individual star's metallicity to obtain precise distances. Including the metallicity term is also important for adopting \rrl{} as a new anchor for the distance scale: the reduced scatter in the MIR PLZ enables the determination of distances with accuracy comparable to Cepheids, fulfilling the promise of an independent route to the cosmological distance scale based on these Population II pulsators.

%


\begin{acknowledgement} 
\noindent\vskip 0.2cm
\noindent {\em Acknowledgments}: This work is based in part on observations made with the Spitzer Space Telescope, which is operated by the JPL-Caltech under a contract with NASA. This work also made use of data from the ESA mission {\it Gaia}, processed by the {\it Gaia} Data Processing and Analysis Consortium (DPAC).
\end{acknowledgement}


\begin{thebibliography}{}
%
%



\bibitem{2016ApJ...832..210B} Beaton, R.~L., Freedman, W.~L., Madore, B.~F., et al., ApJ, \textbf{832},210 (2016)

\bibitem{2016MmSAIt.XXX..XXXB} Bono, G., Braga, V. F., Pietrinferni, A., et al., Mem. SAIt, accepted (2016)

\bibitem{2015ApJ...799..165B} Braga, V.~F., Dall'Ora, M., Bono, G., et al., ApJ, \textbf{799},165 (2015) 

\bibitem{2004ApJ...610..269D} Dall'Ora, M., Storm, J., Bono, G., et al., ApJ, \textbf{610}, 269 (2004) 

\bibitem{2004ApJS..154...10F} Fazio, G.~G., Hora, J.~L., Allen, L.~E., et al., ApJS, \textbf{154},10 (2004) 

\bibitem{1998PhRvD..57.3873F} Feldman, G.~J., Cousins, R.~D., PhRvD, \textbf{57}, 3873 (1998) 

\bibitem{2016A&A...595A...2G} Gaia Collaboration, Brown, A.~G.~A., Vallenari, A., et al., A\&A, \textbf{595}, A2 (2016) 

\bibitem{2012AJ....144...25H} Hendricks, B., Stetson, P.~B., VandenBerg, D.~A., Dall'Ora, M., AJ, \textbf{144},25 (2012) 

\bibitem{2011ApJ...738..185K} Klein, C.~R., Richards, J.~W., Butler, N.~R., Bloom, J.~S., ApJ, \textbf{738}, 185 (2011) 

\bibitem{2013ApJ...776..135M} Madore, B.~F., Hoffman, D., Freedman, W.~L., et al., ApJ, \textbf{776}, 135 (2013) 

\bibitem{2015ApJ...808...50M} Marconi, M., Coppola, G., Bono, G., et al., ApJ, \textbf{808}, 50 (2015)

\bibitem{2017ApJ...XXX.....N} Neeley, J. R., Marengo, M., Bono, G., et al., ApJ, submitted (2017)

\bibitem{2016ApJ...826...56R} Riess, A.~G., Macri, L.~M., Hoffmann, S.~L., et al., ApJ, \textbf{826},56 (2016) 












\end{thebibliography}
%
%

\end{document}